\documentclass[iop,apjl]{emulateapj}

\usepackage{graphicx} 
\usepackage{dcolumn} 
\usepackage{bm} 
\usepackage{amsfonts,amsmath,amssymb,mathrsfs}
\usepackage{color}
\usepackage{epsfig}
\usepackage{natbib,times}
\usepackage{url}
\usepackage{times}

\shorttitle{Progenitors of Short Gamma-Ray Bursts}
\shortauthors{Giacomazzo et al.}

\begin{document}

\title{Compact Binary Progenitors of Short Gamma-Ray Bursts}

\author{Bruno {Giacomazzo}\altaffilmark{1}, Rosalba
  {Perna}\altaffilmark{2}, Luciano {Rezzolla}\altaffilmark{3},
  Eleonora Troja\altaffilmark{4,5}, and Davide
  Lazzati\altaffilmark{6}}

\altaffiltext{1}{JILA, University of Colorado and National Institute
  of Standards and Technology, Boulder, CO 80309, USA}

\altaffiltext{2}{JILA and Department of Astrophysical and Planetary
  Sciences, University of Colorado, Boulder, CO 80309, USA}

\altaffiltext{3}{Max-Planck-Institut f\"ur Gravitationsphysik,
  Albert-Einstein-Institut, Potsdam, D-14476, Germany}

\altaffiltext{4}{NASA Goddard Space Flight Center, Greenbelt, MD
  20771, USA}

\altaffiltext{5}{Department of Astronomy, University of Maryland,
  College Park, MD 20742, USA}

\altaffiltext{6}{Department of Physics, NC State University, 2401
  Stinson Drive, Raleigh, NC 27695-8202, USA}

\begin{abstract}
In recent years, detailed observations and accurate numerical
simulations have provided support to the idea that mergers of compact
binaries containing either two neutron stars (NSs) or an NS and a
black hole (BH) may constitute the central engine of short gamma-ray
bursts (SGRBs). The merger of such compact binaries is expected to
lead to the production of a spinning BH surrounded by an accreting
torus. Several mechanisms can extract energy from this system and
power the SGRBs. Here we connect observations and numerical
simulations of compact binary mergers, and use the current sample of
SGRBs with measured energies to constrain the mass of their powering
tori. By comparing the masses of the tori with the results of fully
general-relativistic simulations, we are able to infer the properties
of the binary progenitors which yield SGRBs. By assuming a constant
efficiency in converting torus mass into jet energy, $\epsilon_{\rm
  jet}=10\%$, we find that most of the tori have masses smaller than
$0.01M_{\odot}$, favoring ``high-mass'' binary NSs mergers, i.e.,
binaries with total masses $\gtrsim1.5$ the maximum mass of an
isolated NS. This has important consequences for the
gravitational-wave signals that may be detected in association with
SGRBs, since ``high-mass'' systems do not form a long-lived
hypermassive NS after the merger. While NS-BH systems cannot be
excluded to be the engine of at least some of the SGRBs, the BH would
need to have an initial spin of $\sim0.9$, or higher.

\end{abstract}


\section{Introduction}
Binary neutron star (BNS) and neutron star-black hole (NS-BH) binaries
are the leading candidates for the central engine of short gamma-ray
bursts (SGRBs; \citealt{1984SvAL...10..177B, 1986ApJ...308L..43P,
  1989Natur.340..126E}). They are also one of the most powerful
sources of gravitational waves (GWs), and advanced interferometric
detectors are expected to observe these sources at rates of
$\sim0.4-400$ and $\sim0.2-300$ events per year for BNS and NS-BH,
respectively~\citep{Abadie:2010}.

Fully general-relativistic simulations have shown how such mergers can
lead to the formation of accretion disks around spinning
BHs~\citep{Baiotti08,Etienne2009,Kiuchi2009,Faber2012}. Moreover, when
magnetic fields are present, they can provide one of the mechanisms
necessary to extract energy, and power collimated relativistic
jets~\citep{2011ApJ...732L...6R, Etienne2012b}.

So far, properties of the progenitors of SGRBs have been inferred by
studying their redshift distribution, close environment, host galaxy
types~\citep{bloom2006, 2009ApJ...703.1696Z, berger2011}, and by
comparing those observations with predictions from population
synthesis models~\citep{PB02,Bel06,oshau08}. In this Letter we make a
connection between theory and observations, which allows us to
directly probe the SGRB progenitors. In particular, we consider a
complete (to date) sample of SGRBs with measured redshifts, and link
the properties of their observed emissions to the masses of the tori
responsible for their generation. By comparing these tori with the
theoretical predictions of~\citet{2010CQGra..27k4105R} and
of~\citet{foucart2012}, we are able to infer the properties of the
compact binaries that may have generated such bursts.

In Section~\ref{grbdata} we provide details on the sample of SGRBs
considered in this Letter. In Section~\ref{tori}, we use the
theoretical results to compute the masses of the tori that have
generated such bursts, and link them to their progenitors. In
Section~\ref{GWs} we show how GWs may be used to further constrain the
progenitors, and in Section~\ref{conclusions} we summarize our main
results.

\section{GRB sample data}
\label{grbdata}

\begin{table}[t!h!]
  \begin{center}
  \caption{SGRB Sample\label{GRBtable}}
  \begin{tabular}{lcccc}
    \tableline\tableline 
    {GRB Name} &
    {$z$} &
    {$E_{\gamma,\rm iso}$ (erg)} &
    $\Delta E$ (keV) &
    {$M_{\rm torus}$ ($M_{\odot}$)}\\
    \tableline
    050509B        & $0.225$  & $9.1 \times 10^{47}$    & $15-150$   & $1.0 \times 10^{-5}$ \\
    050709(EE)     & $0.161$  & $3.4 \times 10^{49}$    & $10-10^4$  & $3.8 \times 10^{-4}$ \\
    050724(EE)     & $0.257$  & $1.9 \times 10^{50}$    & $15-150$   & $2.1 \times 10^{-3}$ \\
    051221A        & $0.546$  & $2.9 \times 10^{51}$    & $10-10^4$  & $3.3 \times 10^{-2}$ \\
    061006(EE)     & $0.438$  & $2.1 \times 10^{51}$    & $10-10^4$  & $2.4 \times 10^{-2}$ \\
    070429B        & $0.902$  & $2.1 \times 10^{50}$    & $15-150$   & $2.3 \times 10^{-3}$ \\
    070714B(EE)    & $0.923$  & $1.6 \times 10^{52}$    & $10-10^4$  & $1.8 \times 10^{-1}$ \\
    071227(EE)     & $0.381$  & $1.2 \times 10^{51}$    & $10-10^4$  & $1.4 \times 10^{-2}$ \\
    080905A        & $0.122$  & $4.5 \times 10^{49}$    & $10-10^4$  & $5.1 \times 10^{-4}$ \\
    090510         & $0.903$  & $4.7 \times 10^{52}$    & $10-10^4$  & $5.2 \times 10^{-1}$ \\
    100117A        & $0.920$  & $1.4 \times 10^{51}$    & $10-10^4$  & $1.6 \times 10^{-2}$ \\
    111117A        & $1.3$    & $5.3 \times 10^{51}$    & $10-10^4$  & $6.0 \times 10^{-2}$ \\
    \tableline                   
    051210         & $1.3$    & $4.0 \times 10^{50}$    & $15-150$     & $4.5 \times 10^{-3}$ \\
    060801         & $1.130$  & $1.9 \times 10^{50}$    & $15-150$     & $2.1 \times 10^{-3}$ \\
    061210(EE)     & $0.410$  & $5.6 \times 10^{50}$    & $15-150$     & $6.2 \times 10^{-3}$ \\
    070724A        & $0.457$  & $2.3 \times 10^{49}$    & $15-150$     & $2.5 \times 10^{-4}$ \\
    070729         & $0.8$    & $1.6 \times 10^{50}$    & $15-150$     & $1.8 \times 10^{-3}$ \\
    080123(EE)     & $0.495$  & $5.7 \times 10^{50}$    & $15-150$     & $6.3 \times 10^{-3}$ \\
    101219A        & $0.718$  & $7.4 \times 10^{51}$    & $10-10^4$    & $8.2 \times 10^{-2}$ \\
    \tableline                     
    060502B    & $0.287$  & $9.8 \times 10^{48}$    & $15-150$     & $1.1 \times 10^{-4}$ \\
    061217     & $0.827$  & $6.8 \times 10^{49}$    & $15-150$     & $7.6 \times 10^{-4}$ \\
    061201     & $0.111$  & $9.4 \times 10^{48}$    & $15-150$     & $1.1 \times 10^{-4}$ \\
    070809     & $0.473$  & $7.9 \times 10^{49}$    & $15-150$     & $8.8 \times 10^{-4}$ \\
    090515     & $0.403$  & $1.0 \times 10^{49}$    & $15-150$     & $1.2 \times 10^{-4}$ \\
    \tableline                                                                                          
  \end{tabular}
  \tablecomments{The different columns refer respectively to the GRB
    name, the redshift $z$ derived from the GRB host, the isotropic
    equivalent gamma-ray energy $E_{\gamma,\rm iso}$, measured in the
    rest-frame energy band $\Delta E$, and the mass of the torus
    $M_{\rm torus}$ (see Equation~(\ref{masstorus})). The different
    blocks refer to the uncertainty in the SGRB/host galaxy
    association~\citep{bloom2002}. The top one includes SGRBs with a
    precise identification of a host galaxy; those in the middle have
    a less certain association with their host; those in the bottom
    are significantly offset from the associated host galaxy, and are
    affected by a larger uncertainty.}
  \end{center}
\end{table}

We selected our sample of SGRBs based on three criteria: duration,
hardness ratio, and spectral lags. {\it Swift} SGRBs with known
redshift are listed in Table~\ref{GRBtable}. SGRBs with a temporally
extended emission (EE) were also considered. In the latter case, the
quoted energetics include the contribution of the short-hard spike,
and of the EE. Since the two emission episodes typically have a
comparable energy budget~\citep{norris11}, the presence of EE affects
our calculations by a factor $\approx2$.

The burst energetics, $E_{\gamma,\rm iso}$ (Column 3 in
Table~\ref{GRBtable}), were calculated by using the prompt emission
spectral parameters (mainly from~\citealt{batcat}
and~\citealt{gbmcat}) and shifted to a common rest-frame energy
band~\citep{bloom01}.  When possible, we used measurements of the
broadband GRB spectrum (e.g., by the {\it Fermi}/GBM) and calculated
$E_{\gamma,\rm iso}$ in the comoving 10~keV-10~MeV energy range. In
most cases, only {\it Swift}/BAT observations are available, and we
report the burst energetics in the narrower 15-150 keV rest-frame
band, thus unavoidably underestimating the bolometric energy release.
For a typical Band spectrum~\citep{band93}, peaking at
$\approx500$~keV~\citep{nava11}, we estimate an average $k$-correction
factor of $\approx6$. We therefore do not expect that the uncertainty
in the GRB spectral shape of {\it Swift} bursts may have a major
impact on the results.

Table~\ref{GRBtable} shows that SGRBs display a wide range of
energies, from $10^{48}$\,erg to $10^{52}$\,erg, with a median value
of $2\times10^{50}$\,erg. The quoted values refer to the isotropic
equivalent gamma-ray energy, while the true energy scale also depends
on the outflow beaming factor $f_b\equiv1-\cos(\theta_{\rm jet})$,
being $\theta_{\rm jet}$ the jet opening angle. The degree of
collimation of SGRBs is still a poorly constrained quantity, inferred
values range from $f_b\approx0.001$ to $f_b\approx0.1$
\citep{burrows06,nicuesa12}, but in most cases only weak lower bounds
can be placed. Here, we use the isotropic energies listed in
Table~\ref{GRBtable} to set an upper limit to the burst-energy
release.

\begin{table}[h!t!]
  \begin{center}
    \caption{BNS Simulations and Torus Masses\label{bnstori_table}}
    \begin{tabular}[c]{lccccc}
      \tableline\tableline
      Model &
      {$M_{\rm BNS}$} &
      {$q$}  &
      $M_{\rm torus}$ &
      $M_{\rm max}$ &
      $M_{\rm BNS}/M_{\rm max}$\\ 
      & {(${M_{\sun}}$)} & & {(${M_{\sun}}$)} & {(${M_{\sun}}$)} & \\
      \tableline
      \texttt{1.46-45-IF} & $3.24$ & $1.00$ & $0.1374$ & $2.20$ & $1.47$\\
      \texttt{1.62-45-IF} & $3.61$ & $1.00$ & $0.1101$ & $2.20$ & $1.64$\\

      \tableline
      \texttt{M3.6q1.00} & $3.90$ & $1.00$ & $0.0012$ & $2.20$ & $1.77$\\    
      \texttt{M3.7q0.94} & $4.03$ & $0.94$ & $0.0121$ & $2.20$ & $1.83$\\    
      \texttt{M3.4q0.91} & $3.76$ & $0.92$ & $0.1202$ & $2.20$ & $1.71$\\   
      \texttt{M3.4q0.80} & $3.72$ & $0.81$ & $0.2524$ & $2.20$ & $1.69$\\    
      \texttt{M3.5q0.75} & $3.80$ & $0.77$ & $0.1939$ & $2.20$ & $1.73$\\
      \texttt{M3.4q0.70} & $3.71$ & $0.72$ & $0.2558$ & $2.20$ & $1.69$\\

      \tableline
      \texttt{APR145145} & $2.87$ & $1.00$ & $0.000549$ & $2.18$ & $1.32$\\
      \texttt{APR1515}   & $2.97$ & $1.00$ & $0.000134$ & $2.18$ & $1.36$\\
      \texttt{APR1316}   & $2.87$ & $0.81$ & $0.0275$   & $2.18$ & $1.32$\\
      \texttt{APR135165} & $2.97$ & $0.82$ & $0.00707$  & $2.18$ & $1.36$\\

      \tableline
      \texttt{APR4-28} & $2.77$ & $1.00$ & $0.003$  & $2.21$ & $1.25$\\
      \texttt{SLy-27}  & $2.67$ & $1.00$ & $0.02$   & $2.05$ & $1.30$\\
      \texttt{H3-27}   & $2.68$ & $1.00$ & $0.05$   & $1.79$ & $1.50$\\
      \texttt{H3-29}   & $2.87$ & $1.00$ & $0.01$   & $1.79$ & $1.61$\\
      \texttt{H4-27}   & $2.68$ & $1.00$ & $0.18$   & $2.03$ & $1.32$\\
      \texttt{H4-29}   & $2.87$ & $1.00$ & $0.02$   & $2.03$ & $1.41$\\
      \texttt{H4-30}   & $2.97$ & $1.00$ & $0.01$   & $2.03$ & $1.46$\\
      \texttt{ALF2-27} & $2.67$ & $1.00$ & $0.16$   & $2.09$ & $1.28$\\
      \texttt{ALF2-29} & $2.87$ & $1.00$ & $0.02$   & $2.09$ & $1.38$\\
      \texttt{ALF2-30} & $2.97$ & $1.00$ & $0.003$  & $2.09$ & $1.42$\\
      \texttt{PS-27}   & $2.68$ & $1.00$ & $0.04$   & $1.76$ & $1.53$\\
      \texttt{PS-29}   & $2.88$ & $1.00$ & $0.02$   & $1.76$ & $1.64$\\
      \texttt{PS-30}   & $2.97$ & $1.00$ & $0.01$   & $1.76$ & $1.69$\\
      \tableline
    \end{tabular}
    \tablecomments{The different columns represent respectively the
      name of the model, the gravitational mass of the binary, $M_{\rm
        BNS}$, the mass ratio of the gravitational masses of the two
      NSs, $q$, the baryonic mass of the torus, $M_{\rm torus}$, the
      maximum gravitational mass of an isolated NS for the equation of
      state (EOS) used in that simulation, $M_{\rm max}$, and the
      ratio between the mass of the binary and $M_{\rm max}$, $M_{\rm
        BNS}/M_{\rm max}$. The different blocks of the table refer,
      from top to bottom, to the simulations by~\cite{Baiotti08},
      \cite{2010CQGra..27k4105R}, \cite{Kiuchi2009},
      and~\cite{Hotokezaka2011}. Note that the simulations reported
      in~\cite{Baiotti08} and~\cite{2010CQGra..27k4105R} used an
      ideal-fluid EOS and hence they can be scaled to different
      masses. Here we have chosen the values for an ideal-fluid EOS so
      that $M_{\rm max}=2.20$, in agreement with current
      observations~\citep{Demorest2010}.}
  \end{center}
\end{table}

Assuming that all the SGRBs in our sample were produced by accretion
tori around spinning BHs, we now correlate the values for the
isotropic energy listed in Table~\ref{GRBtable} with the mass of such
tori. In particular the torus mass is determined as

\begin{figure*}[t!]
  \centering
  \begin{tabular}{cc}
    \includegraphics[height=.35\textwidth]{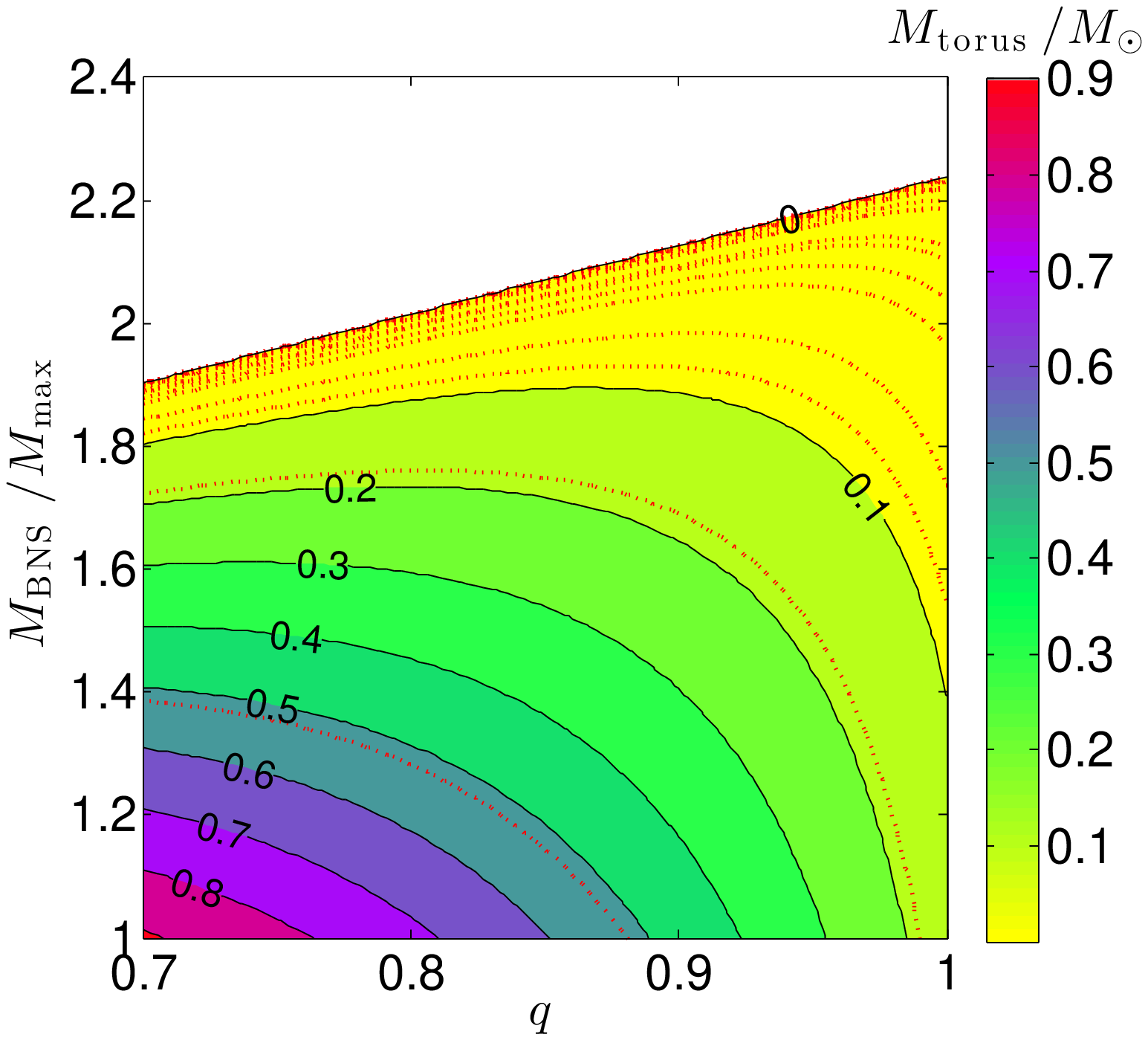}
    \includegraphics[height=.35\textwidth]{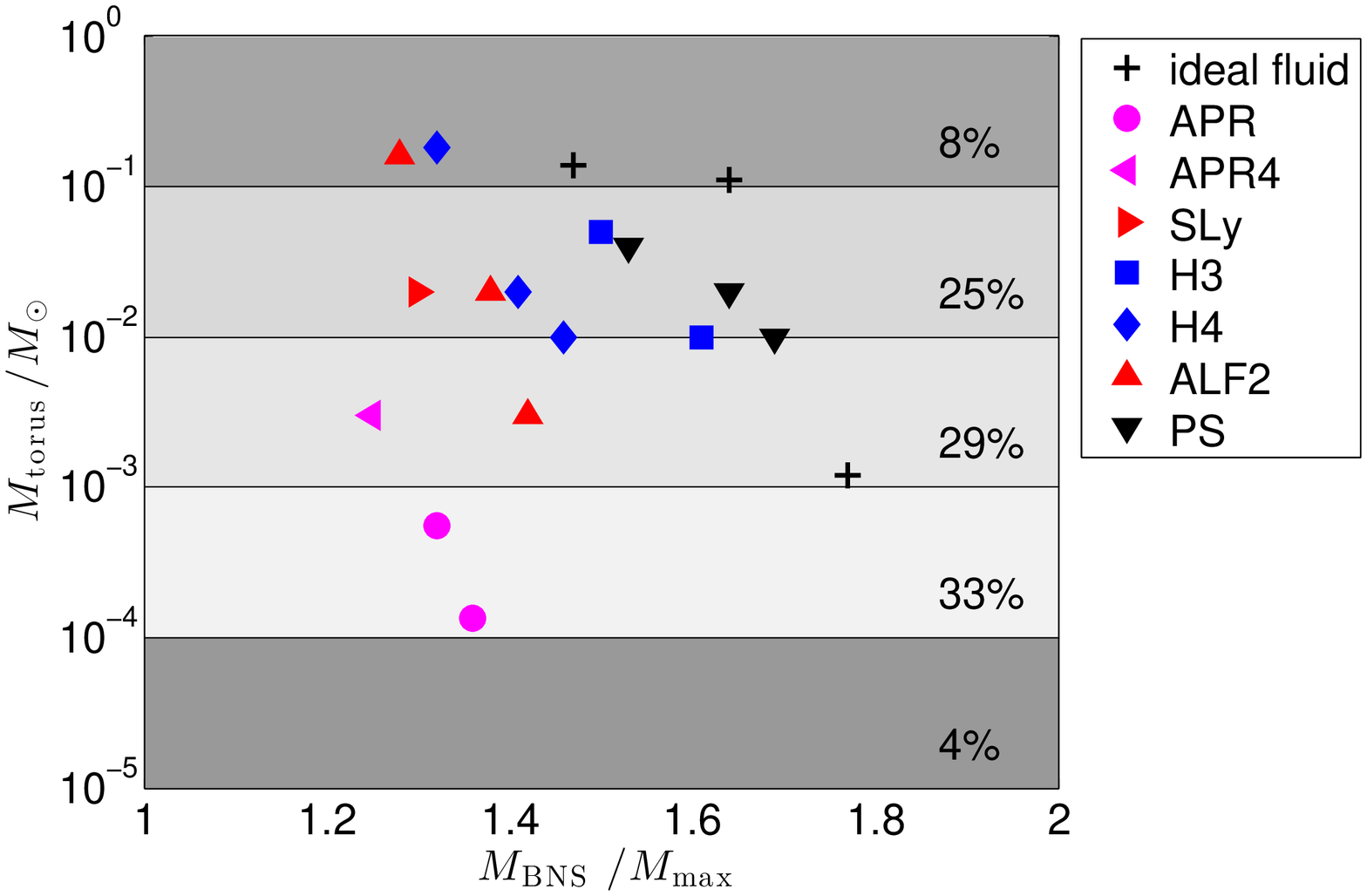}
  \end{tabular}
  \caption{Left panel: $M_{\rm torus}/M_\odot$ as a function of the
    mass ratio $q$ and of the ratio between the gravitational mass of
    the binary and the maximum mass for an isolated NS ($M_{\rm
      BNS}/M_{\rm max}$). $M_{\rm torus}$ has been computed using
    Equation~(\ref{eqmass}). The dotted lines are the isocontours
    corresponding to the $M_{\rm torus}$ values in
    Table~\ref{GRBtable}. Right panel: plot of $M_{\rm torus}$ as a
    function of $M_{\rm BNS}/M_{\rm max}$ for all the equal-mass
    ($q=1$) simulations reported in Table~\ref{bnstori_table}. The
    horizontal bars give the percentage of the SGRBs in
    Table~\ref{GRBtable} that are generated by tori with that range of
    masses (assuming a total efficiency $\epsilon$ of 5\% as in
    Table~\ref{GRBtable}). Since the mass of the torus increases for
    $q<1$ , each point should be considered as a lower limit on the
    mass that can be obtained for that EOS and mass of the binary. The
    different points refer to $M_{\rm torus}$ computed from
    simulations of BNS mergers using different EOSs (see
    Table~\ref{bnstori_table}).\label{figure1}}
\end{figure*}

\begin{equation}
E_{\gamma,\rm iso}=\epsilon M_{\rm torus}c^{2} \,, \label{masstorus}
\end{equation}
where $\epsilon$ is the efficiency in converting the mass of the torus
$M_{\rm torus}$ into the isotropic gamma-ray emission
$E_{\gamma,\rm iso}$. Here, $\epsilon$ is given by the product of two
efficiencies: one to convert mass of the torus into jet energy,
$\epsilon_{\rm jet}$, and the other to convert the latter into
gamma-rays, $\epsilon_\gamma$.

Fully general-relativistic simulations of BNS mergers have shown the
formation of thin and highly magnetized tori around spinning
BHs~\citep{2010CQGra..27k4105R,2011ApJ...732L...6R}. Here we make the
important assumption that SGRBs are powered via magnetic
fields~\citep{BZ77,BP82,Tchekhovskoy2012} and ignore the effects of
viscosity and neutrino cooling~\citep{chen2007}.\footnote{For a
  discussion of neutrino-powered SGRBs and the relation between
  simulations and observations, see~\citet{Lee2005},
  \citet{Oechslin2006}, and \citet{fan-wei}.} General-relativistic
magnetohydrodynamic (GRMHD) simulations of accretion disks showed that
the efficiency in converting torus mass and BH spin into jet energy
(i.e., $\epsilon_{\rm jet}$) varies between few per cent up to more
than 100\% for maximally spinning
BHs~\citep{devilliers2005,Tchekhovskoy2011,McKinney2012,Fragile2012}. The
efficiency depends sensitively on the BH spin, the disk thickness, and
the magnetic flux. Accounting for all of these effects is currently
not possible, and thus we made the simplifying assumption of a
constant efficiency for all BNS and NS-BH mergers, $\epsilon_{\rm
  jet}=10\%$. While an obvious approximation, our main results do not
change sensitively if $\epsilon_{\rm jet}$ is taken to be larger than
$\sim$0.1\%.

After the jet is emitted, a fraction of its energy is converted into
gamma rays. The conversion efficiency for a sample of long and short
{\it Swift} GRBs was computed by \citet{zhang07} by comparing the
gamma-ray fluence with the brightness of the X-ray afterglow at early
and late times. They find that, while the efficiency in long GRBs
varies strongly from burst to burst, ranging from a fraction of a per
cent to almost 100\%, in SGRBs the range is narrower, varying between
30\% and 60\%, with an average of 49\%. We hence assume a fiducial
value of $\epsilon_\gamma=50\%$, so that the total efficiency in
Equation~(\ref{masstorus}) becomes $\epsilon=5\%$. The last column of
Table~\ref{GRBtable} shows the corresponding torus masses.

\section{Torus masses}
\label{tori}
In the following, we link $M_{\rm torus}$ to the theoretical
predictions of~\citet{2010CQGra..27k4105R} and of~\citet{foucart2012},
who derived analytic fits from the results of fully
general-relativistic simulations of BNS and NS-BH mergers,
respectively (see also~\citealt{Pannarale2010}).

\subsection{Binary Neutron Star Mergers}
\label{bnstori}

\citet{2010CQGra..27k4105R} derived a phenomenological expression to
compute the masses of the tori formed by BNS mergers. Here we have
revised that fit and expressed it as a function of two dimensionless
quantities: the gravitational mass ratio $q\leq1$ and the ratio
between the gravitational mass of the binary and the maximum
gravitational mass for an isolated NS ($M_{\rm BNS}/M_{\rm max}$). We
derived
\begin{equation}
    M_{\rm torus}=[c_1 (1-q) + c_2][c_3 (1+q)-M_{\rm BNS}/M_{\rm max}] \,.
    \label{eqmass}
\end{equation}
The coefficients $c_1=2.974\pm3.366$, $c_2=0.11851\pm0.07192$, and
$c_3=1.1193\pm0.1579$ were determined by fitting Equation~(\ref{eqmass}) to
the results of the fully general-relativistic simulations
of~\cite{Baiotti08} and~\cite{2010CQGra..27k4105R}, but rescaled to
allow for a value of $M_{\rm max}=2.20\,M_{\odot}$ to be more
consistent with current observations of NS masses (see also
Table~\ref{bnstori_table}).

The left panel of Figure~\ref{figure1} shows $M_{\rm torus}$ computed
using Equation~(\ref{eqmass}) as a function of $q$ and $M_{\rm
  BNS}/M_{\rm max}$, while each of the red dotted lines represents the
isocontour relative to an observed GRB in Table~\ref{GRBtable} when
assuming our fiducial value $\epsilon=5\%$. The right panel shows the
distribution of torus masses obtained from observations (horizontal
bars, see Table~\ref{GRBtable}) together with the mass of the torus
computed from numerical simulations of equal-mass BNSs ($q=1$, see
Table~\ref{bnstori_table}). As one can easily see, two thirds of the
SGRBs of our sample appear to be generated by tori with masses smaller
than $\sim10^{-2}M_{\odot}$. Moreover, since $M_{\rm torus}$ increases
for $q<1$, each point should be considered as a lower limit on the
mass that can be obtained for that equation of state (EOS) and BNS
mass. This means that, while the energetics of most SGRBs can be
explained by current numerical simulations, some of the less energetic
SGRBs should result from BNS mergers with masses larger than the ones
simulated so far (since $M_{\rm torus}$ decreases with increasing
$M_{\rm BNS}$).

\begin{figure*}[h!t!]
  \centering
  \begin{tabular}{cc}
    \includegraphics[width=.4\textwidth]{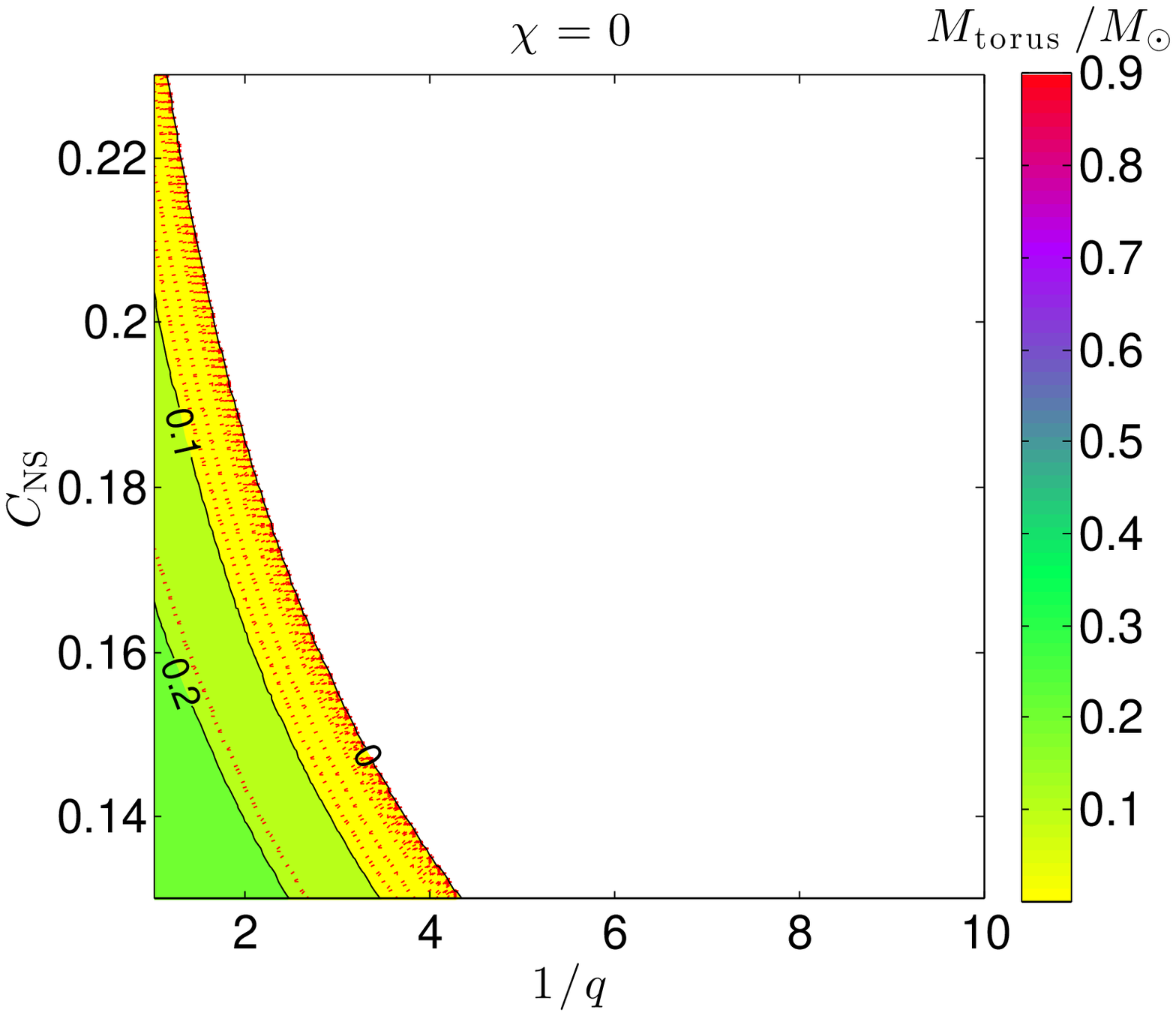}
    \includegraphics[width=.4\textwidth]{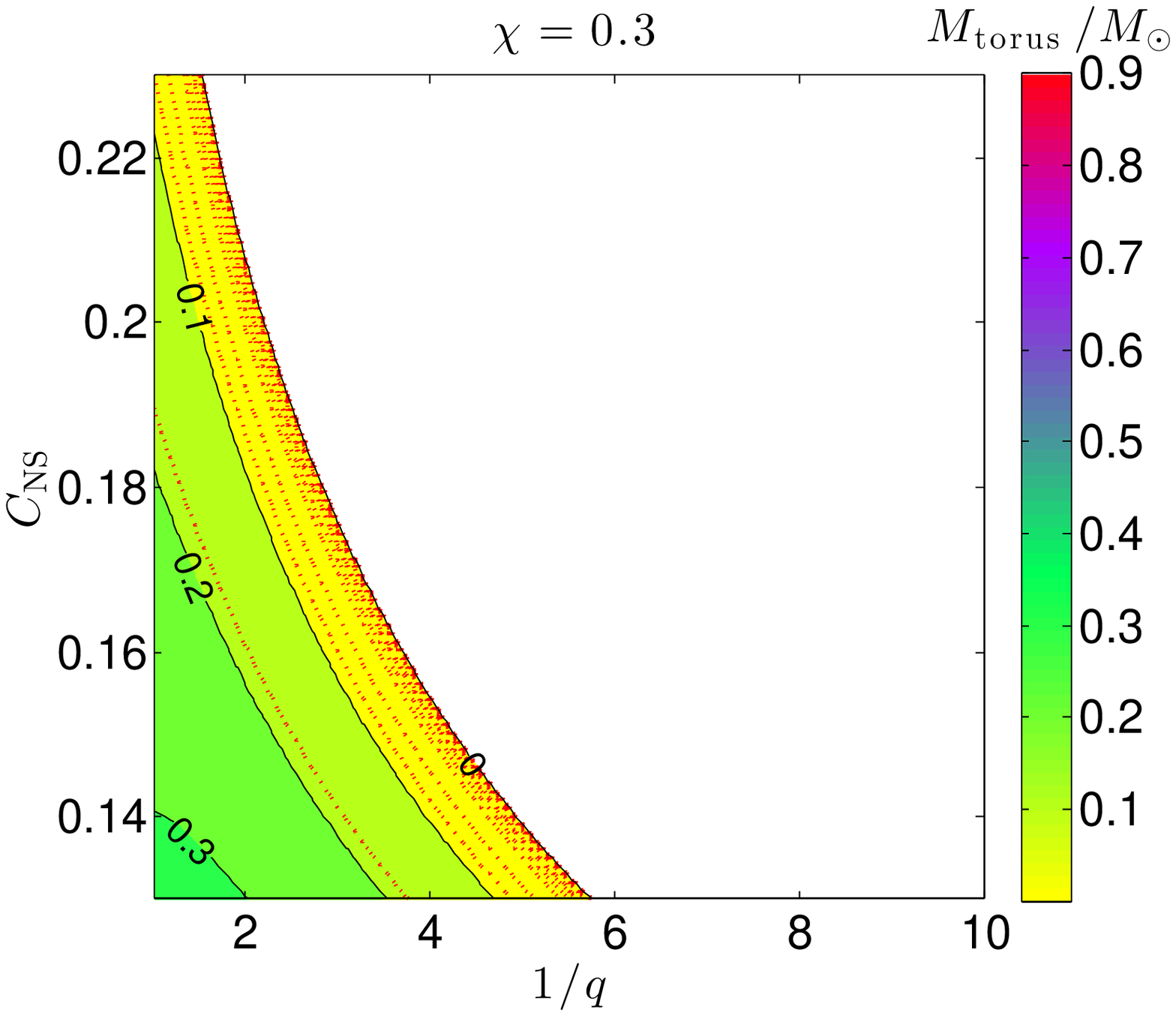}\\
    \includegraphics[width=.4\textwidth]{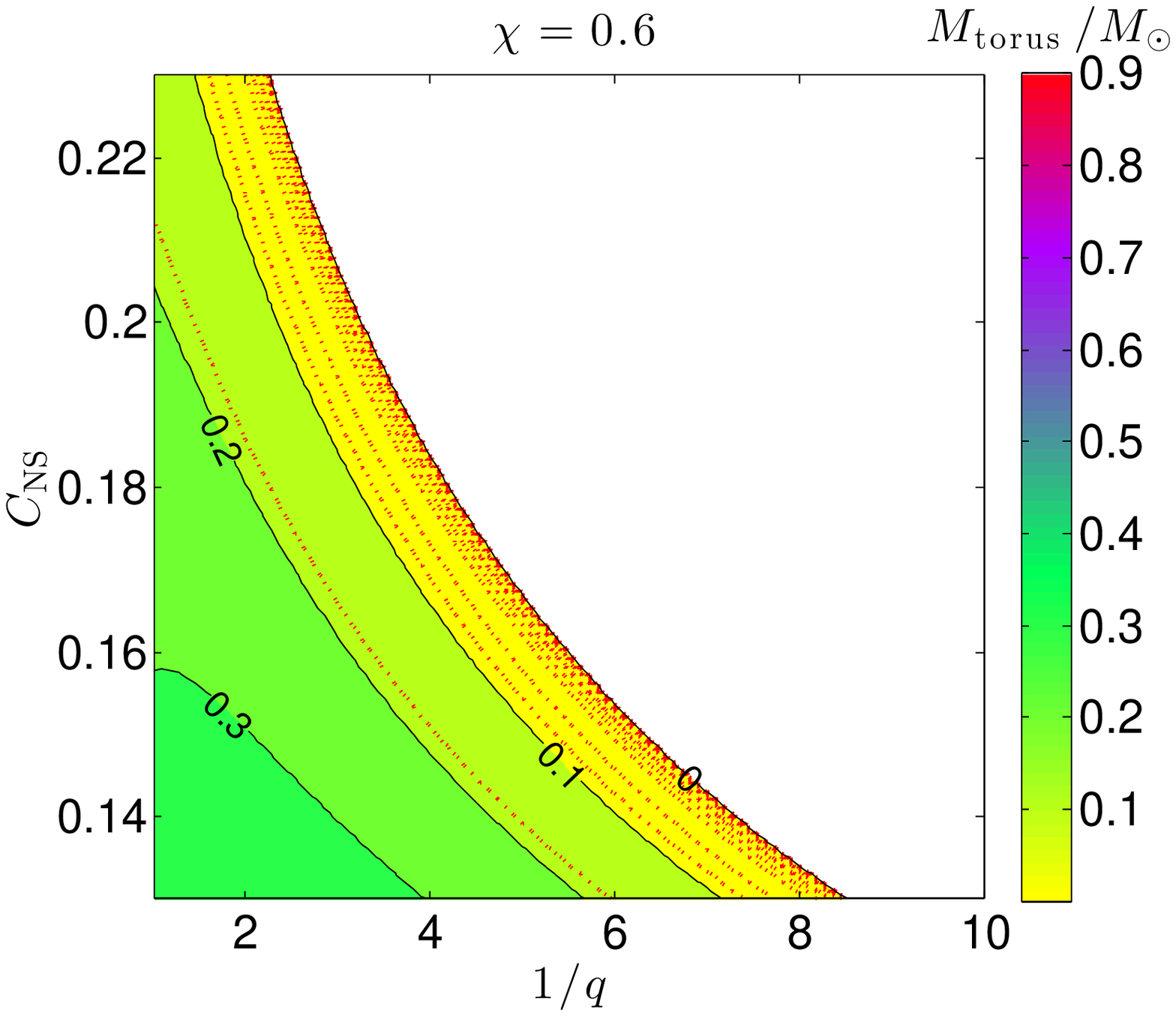}
    \includegraphics[width=.4\textwidth]{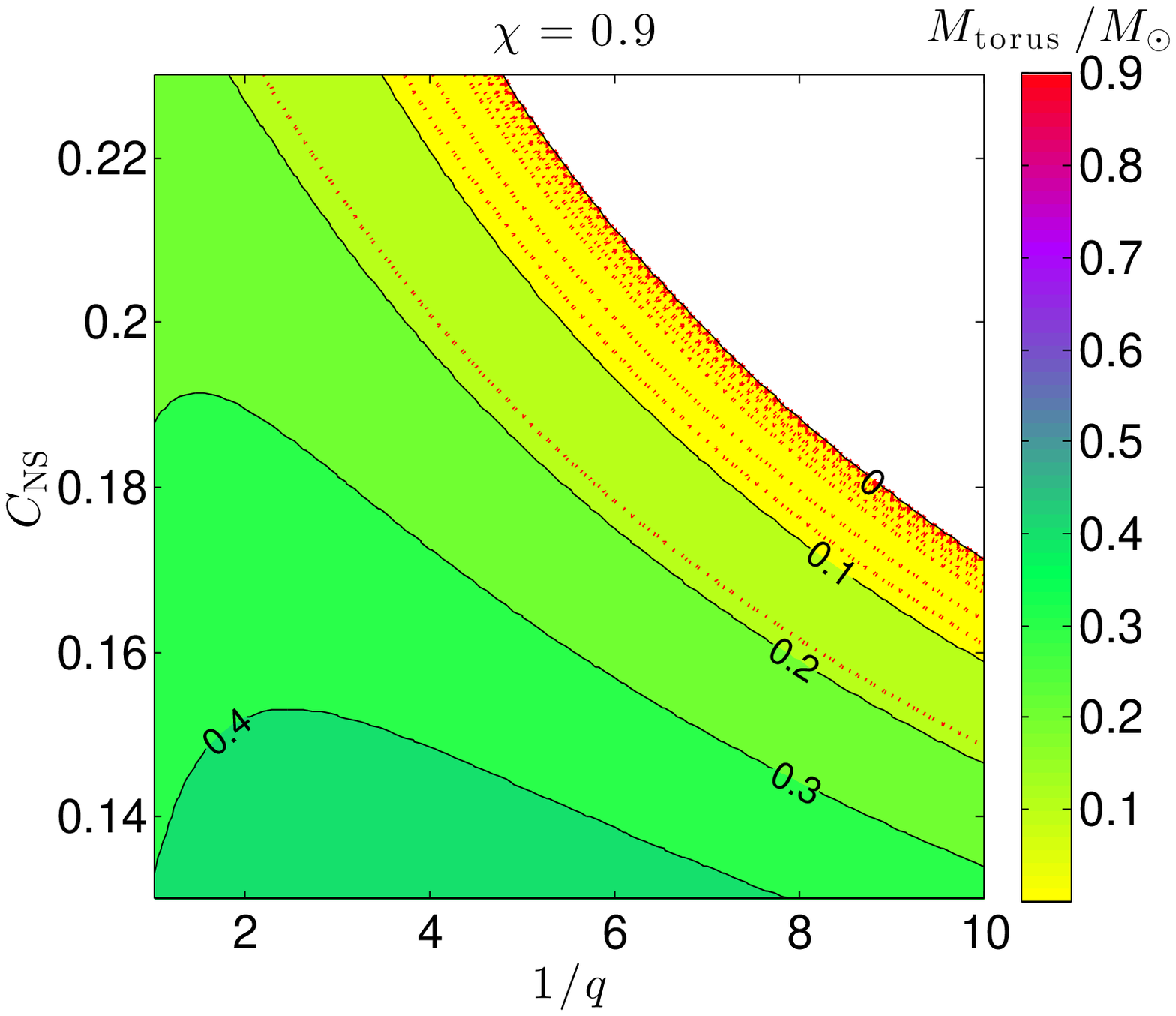}
  \end{tabular}
  \caption{Similar to the left panel of Figure~\ref{figure1}, but for
    the NS-BH case. $M_{\rm torus}$ is shown as a function of the NS
    compactness $C_{\rm NS}$ and mass ratio $1/q\equiv M_{\rm
      BH}/M_{\rm NS}$. $M_{\rm torus}$ is computed using
    Equation~(\ref{fitbhns}). Each panel assumes a different value for
    the dimensionless spin of the BH, $\chi$. In all cases, we assume
    $M_{\rm NS}=1.4\,M_{\sun}$. The dotted lines are the isocontours
    corresponding to the $M_{\rm torus}$ values in
    Table~\ref{GRBtable}.
    \label{figure3}}
\end{figure*}

It is evident from the left panel of Figure~\ref{figure1} that, for an
ideal-fluid EOS, almost all of the SGRBs would be generated by BNSs
with $M_{\rm BNS}/M_{\rm max}>1.8$ and hence they would be
``high-mass'' systems. This means that the mass of the system would be
too high to lead to the formation of a long-lived hypermassive NS
(HMNS) and that the merger would produce a prompt collapse to BH. This
is also true for the models with realistic EOSs shown in the right
panel of Figure~\ref{figure1}. For example, the two circles refer to
simulations of equal-mass binaries using an APR EOS (models
\texttt{APR145145} and \texttt{APR1515}; see~\citealt{Kiuchi2009}) and
they produce tori with masses in the range of $\sim33\%$ of all SGRBs
in our sample. As reported in~\citet{Kiuchi2009}, in both cases,
collapse to BH occurs $\sim1$ ms after merger. We recall that the
threshold $M_{\rm BNS}/M_{\rm max}$ below which a long-lived HMNS is
formed is strongly dependent on the EOS. All the simulations that
produce tori with masses $\la0.1M_{\sun}$ in the right panel of
Figure~\ref{figure1} produce an HMNS that collapses on a timescale of
few ms~\citep{Kiuchi2009, 2010CQGra..27k4105R,
  Hotokezaka2011}.\footnote{If $\epsilon$ was one order of magnitude
  smaller, $M_{\rm torus}$ would be $10$ times larger, but 67\% of the
  SGRBs would still have $M_{\rm torus}<0.1M_\odot$ and hence be
  generated by ``high-mass'' systems.} As we discuss in
Section~\ref{GWs}, this has a fundamental impact on the GW signal we
may expect from SGRBs.

\subsection{NS-BH Mergers}
\label{nsbhtori}

\cite{foucart2012} derived the following fit for the mass of the
torus produced by an NS-BH merger:
\begin{equation}
  M_{\rm torus}=\left( \frac{M^b_{\rm NS}}{M_{\rm NS}}\right) \left[\alpha (3/q)^{1/3} (1-2C_{\rm NS})M_{\rm NS}-\beta R_{\rm isco} C_{\rm NS} \right] \,,
\label{fitbhns}
\end{equation}
where $\alpha=0.288\pm0.011$, $\beta=0.148\pm0.007$, $M_{\rm NS}$ is
the gravitational mass of the NS, $M^b_{\rm NS}$ its baryonic mass,
$C_{\rm NS}$ the NS compactness, $1/q\equiv M_{\rm BH}/M_{\rm NS}>1$
the ratio between the BH and NS masses, and $R_{\rm isco}$ the radius
of the innermost stable circular orbit~\citep{foucart2012}. We note
that in order to compute $M_{\rm torus}$ we need to know the ratio
$M_{\rm NS}/M^b_{\rm NS}$, for which there is no analytic expression
available. We make here the reasonable assumption that the baryonic
mass is 10\% larger than the gravitational mass.\footnote{For the NSs
  reported in Table~\ref{bnstori_table}, the baryonic mass $M^b_{\rm
    NS}$ is $\sim8\%$ larger than the gravitational mass $M_{\rm NS}$
  for the ideal-fluid EOS and $\sim11\%$ larger for the APR EOS.}
This assumption may lead to a few percent error on the mass of the
torus, which is sufficiently small to not affect the results of this
Letter.

The four panels in Figure~\ref{figure3} show $M_{\rm torus}$, computed
using Equation~(\ref{fitbhns}), as a function of the NS compactness
$C_{\rm NS}$ and mass ratio $1/q$. Each panel assumes a different
value for the dimensionless spin of the BH, $\chi$. It is evident from
this figure that not even the most rapidly spinning BH ($\chi=0.9$)
can explain the most energetic burst in our sample (GRB090510 with
$M_{\rm torus}\sim0.5M_{\sun}$). Moreover, if we account for the
results of populations synthesis calculations~\citep{Belczynski2008},
which predict most of the NS-BH binaries to have mass ratios $1/q$
between $\sim7$ and $\sim10$, while the NS compactness is expected to
be larger than $\sim0.16$~\citep{Steiner2012}, then most of the SGRBs
in our sample can be only explained if the binary has a BH with an
initial spin of $\sim0.9$ or larger. From current observations of
SGRBs, it is then clear that, while current simulations of BNS mergers
may easily produce tori in the range required to explain all the
current observations, NS-BH mergers cannot be used to explain the most
energetic bursts.\\

\begin{figure*}[h!t!]
  \centering
  \begin{tabular}{ccc}
    \includegraphics[width=.29\textwidth]{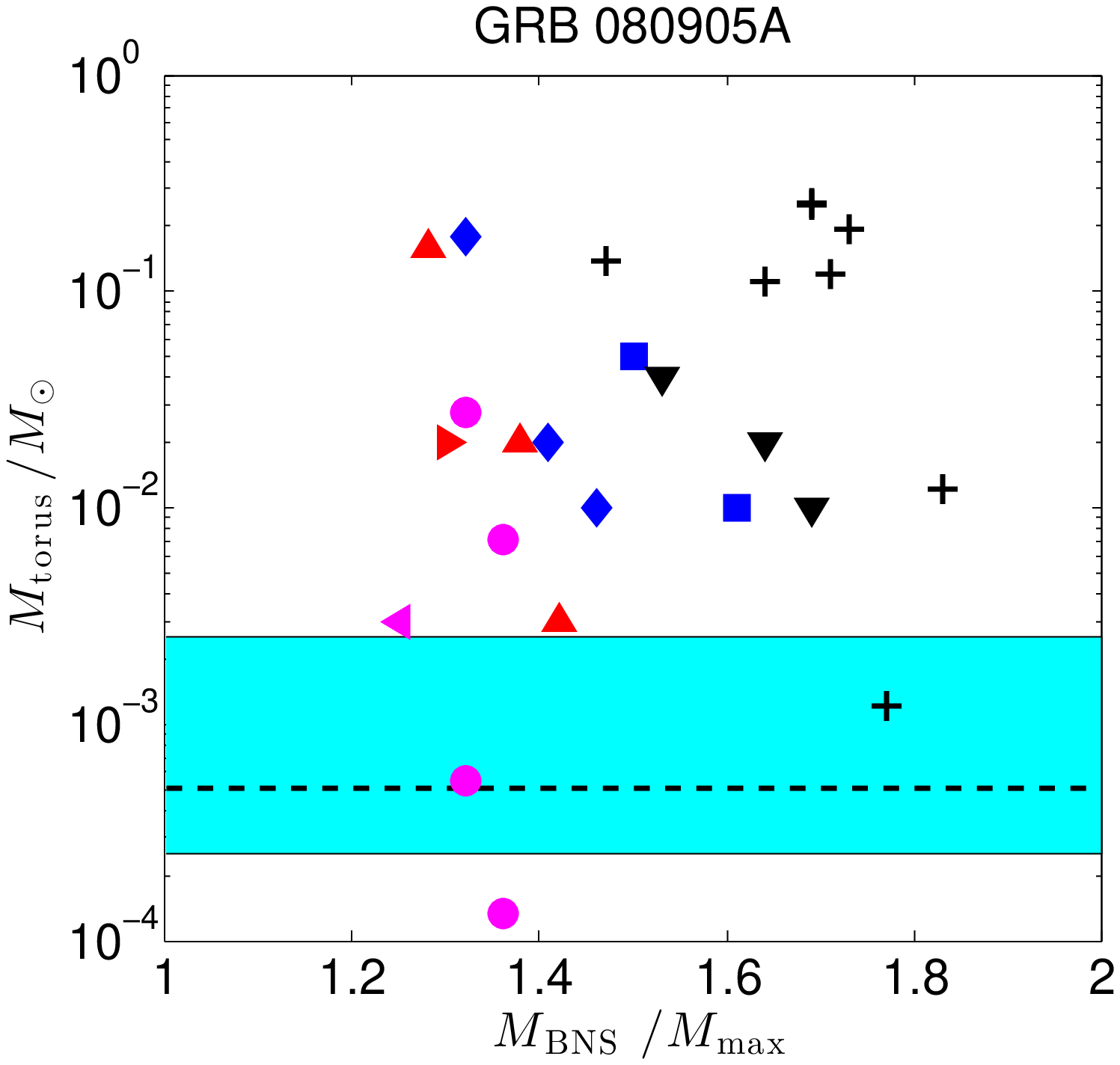}&
    \includegraphics[width=.29\textwidth]{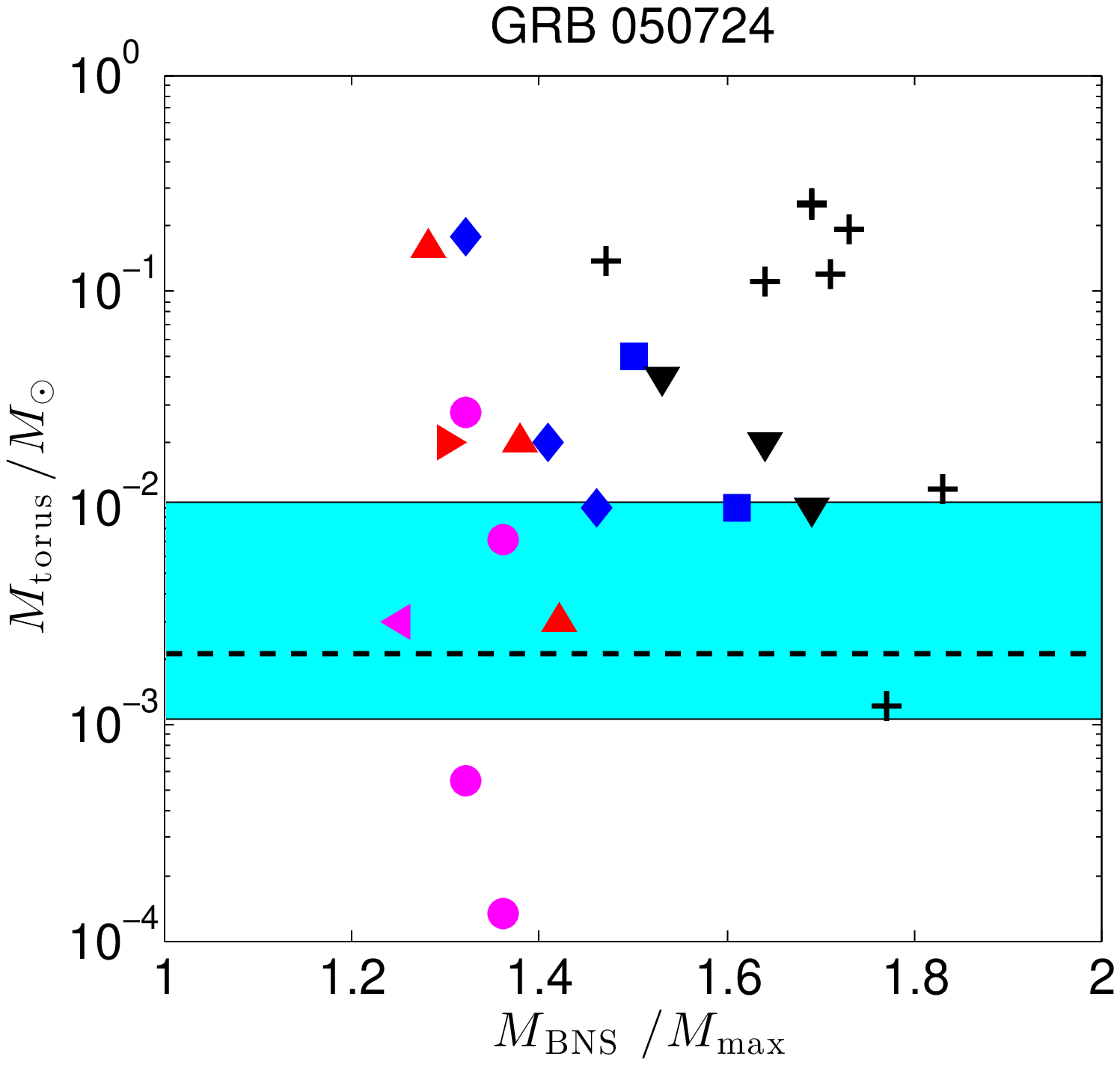}&
    \includegraphics[width=.29\textwidth]{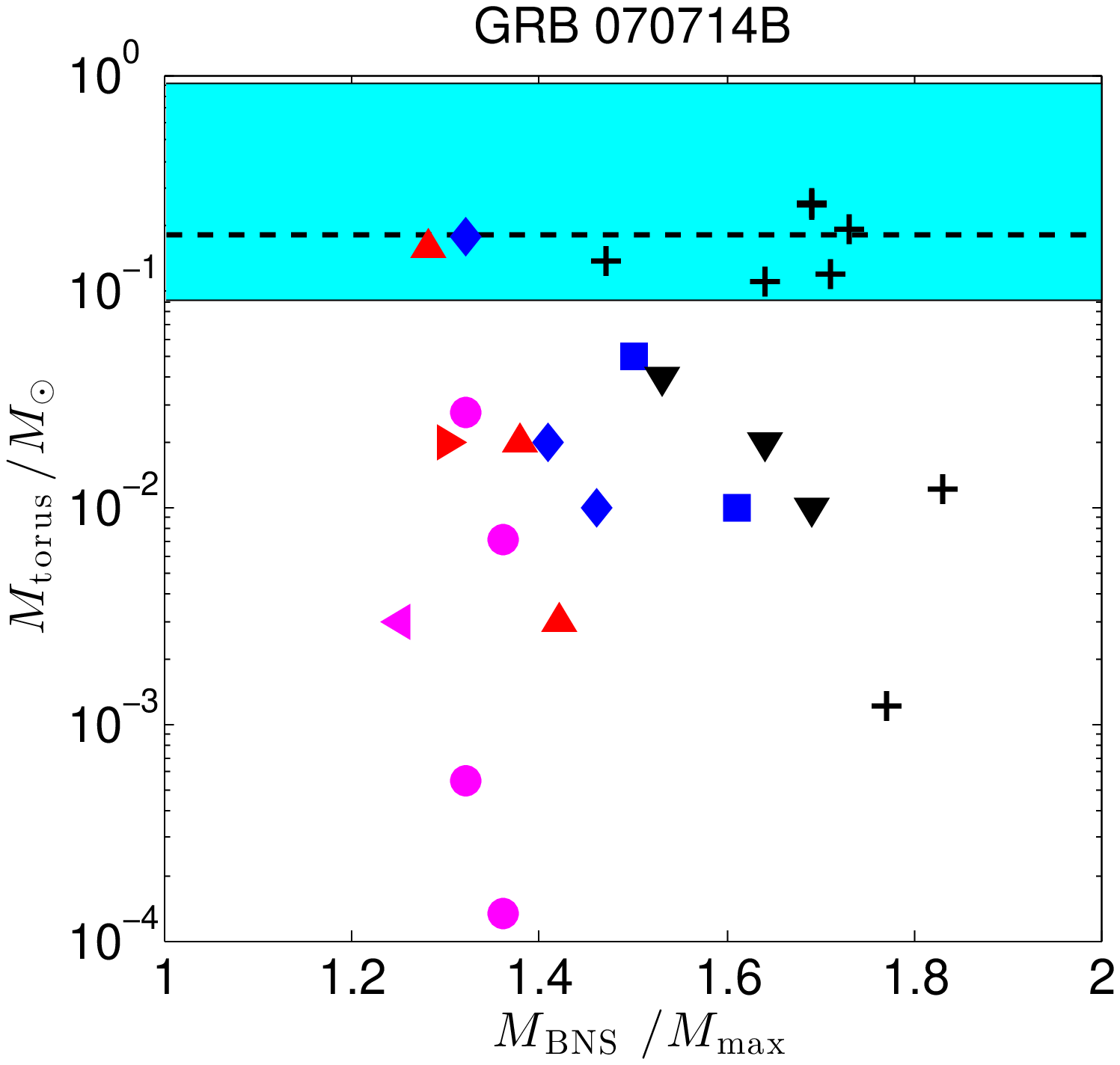}\\
  \end{tabular}
  \caption{Different panels plot $M_{\rm torus}$ as a function of
    $M_{\rm BNS}/M_{\rm max}$ for all the simulations reported in
    Table~\ref{bnstori_table}, and compare them with three SGRBs taken
    from Table~\ref{GRBtable}: GRB080905A, GRB050724, and
    GRB070714B. In each panel, a horizontal dashed line represents the
    value of $M_{\rm torus}$ reported in Table~\ref{GRBtable} while
    the shaded region represents the range of $M_{\rm torus}$ assuming
    a total efficiency between $\epsilon=1\%$ and
    $\epsilon=10\%$. Symbols for the various EOS from
    Table~\ref{bnstori_table} are the same as in Figure~\ref{figure1},
    but here also the $q\neq1$ simulations have been
    included.\label{figure2}}
\end{figure*}

\section{Constraints using future GW observations}
\label{GWs}

As shown in~\citet{Baiotti08} and~\citet{2010CQGra..27k4105R}, the GW
signal is strongly affected by the mass of the system and how close
this is to the maximum mass for each particular EOS. BNSs with masses
close to the maximum mass exhibit a prompt collapse to BH after the
merger, while lower-mass systems produce an HMNS which can survive
from few ms up to hundreds of
ms~\citep{Baiotti08,2010CQGra..27k4105R,Giacomazzo2011}. GW signals
from ``high-mass'' systems are simply composed of the inspiral,
merger, and BH ring-down phases. Lower-mass systems, instead, display
a more complex GW signal with a rich spectrum due to the emission of
GWs from the HMNS formed after the merger. Such emission is important
since it can help infer the properties of the NS
EOS~\citep{2012arXiv1204.1888B}. However, since the GW signal emitted
by the HMNS is in a range of frequencies between $\sim2$kHz and
$\sim4$kHz~\citep{Baiotti08, 2012arXiv1204.1888B}, it may be difficult
for advanced LIGO/Virgo to detect it, and a third generation of
detectors, such as the Einstein Telescope, would be
required~\citep{Andersson:2009yt}. On the other hand, the formation of
an HMNS after the merger can also be inferred by measuring the delay
time between the BNS merger (indicated by the GW signal) and the time
of the emission of the SGRB (which we may assume coincident with BH
formation). A delay time of $\sim100$ ms or larger would clearly
indicate the formation of an HMNS.

As discussed in Section~\ref{bnstori}, we find that only the most
energetic SGRBs can be compatible with a low-mass binary and hence the
formation of an HMNS. For the greatest majority of SGRBs, a high-mass
system is the most likely scenario, and hence we expect SGRBs to be
observed simultaneously with GWs which would lack the high-frequency
emission typical of the HMNS. Although the GW signal from a prompt
collapse is not as rich as that from an HMNS, the simultaneous
detection of an SGRB with the associated GW may help considerably in
constraining the NS EOS.

This is illustrated in Figure~\ref{figure2} with a selection of three
SGRBs from our sample (with low, medium, and high energetics).  In
each panel, the horizontal dashed line represents the value of $M_{\rm
  torus}$ reported in Table~\ref{GRBtable}, while the shaded region
represents the range of $M_{\rm torus}$ assuming a total efficiency
$\epsilon$ between 1\% and 10\%. The various points represent $M_{\rm
  torus}$ computed from simulations of BNS mergers using different
EOSs (see Table~\ref{bnstori_table}). If an SGRB was detected together
with a GW signal, we could use the energetic of the burst to determine
the torus mass (horizontal bars in Figure~\ref{figure2}), while the GW
signal could be used to infer the mass of the BNSs (which would give a
vertical bar in those panels). The combination of these two pieces of
information would restrict the allowed EOS parameter space.

In the case of NS-BH binaries, the mass ratio $q$ and the spin of the
BH can in principle be measured via GW observations; then a
simultaneous detection of a GW and an SGRB would allow to set an
independent constraint on the NS compactness $C_{\rm NS}$ and hence
infer the NS EOS.

\section{Summary}
\label{conclusions}
We have performed a novel analysis of the energetics of SGRBs in
connection with the properties of the compact binary systems that may
have generated them. We have shown that most of the SGRBs could be
produced by magnetized tori with masses lower than $\sim 0.01
M_{\sun}$. Combining this information with the results of numerical
simulations of NS-NS and NS-BH mergers, we have concluded that most of
the SGRBs are consistent with the merger of ``high-mass'' BNS systems,
i.e., with $M_{\rm BNS}/M_{\rm max}\gtrsim1.5$. While NS-BH systems
cannot be completely excluded, the BH would need to have an initial
spin of $\sim0.9$, or higher. Moreover, the most energetic SGRBs, such
as GRB090510, could not be produced by the merger of an NS with a
BH.\footnote{See~\citet{McWilliams2011} for an alternative.} We note
that while our results are affected by some uncertainty in the exact
value of the efficiency $\epsilon$, our conclusions are robust as long
as $\epsilon_{\rm jet}\gtrsim0.1\%$ (i.e.,
$\epsilon\gtrsim5\times10^{-4}$), which is much lower than what was
observed in GRMHD simulations of jets from accretion
disks~\citep{devilliers2005,Tchekhovskoy2011,McKinney2012,Fragile2012}.

GW signals from SGRBs would help validate our results.  In particular,
in the case of BNSs, since we find that SGRBs are most likely
generated by ``high-mass'' BNSs, the GW signal would lack the features
that are associated with the formation of an HMNS, since ``high-mass''
BNSs produce a prompt collapse to BH a few ms after merger. 
A simultaneous detection of GWs with SGRBs would help constrain
the EOS of NS matter.

\acknowledgments We thank J.~M. Demopoulos, H.-T. Janka, and
S.~T. McWilliams for useful comments. B.G. and R.P. acknowledge
support from NSF Grant No. AST 1009396 and NASA Grant No. NNX12AO67G.
L.R.  acknowledges support from the DFG grant SFB/Transregio~7 and by
``CompStar'', a Research Networking Programme of the ESF.

\bibliographystyle{apj}

\end{document}